\documentclass[aps,prd,floats,twocolumn,epsf,nofootinbib,showpacs]{revtex4}
\usepackage{latexsym}
\usepackage{amssymb}
\usepackage[dvips]{graphicx}
\usepackage[applemac]{inputenc}
\usepackage[nottoc]{tocbibind} 
\usepackage{float}
\usepackage{fancyhdr}
\usepackage{amsfonts}
\usepackage{amsmath}
\usepackage{color}
\usepackage{mathrsfs}
\usepackage{bm}
\usepackage{epsfig}

\newcommand{\be}{\begin{equation}}
\newcommand{\ee}{\end{equation}}

\def\lsim{\mathrel{\raise.3ex\hbox{$<$\kern-.75em\lower1ex\hbox{$\sim$}}}}

\def\gsim{\mathrel{\raise.3ex\hbox{$>$\kern-.75em\lower1ex\hbox{$\sim$}}}}

\def\beq{\begin{eqnarray}}
\def\eeq{\end{eqnarray}}
\def\bea{\begin{eqnarray}}
\def\eea{\end{eqnarray}}

\begin{document}

\title{Toward A Consistent Picture For CRESST, CoGeNT and DAMA}

\author{Chris Kelso$^{a,b}$, Dan Hooper$^{a,c}$ and Matthew R. Buckley$^{a}$}

\affiliation{
$^a$Center for Particle Astrophysics, Fermi National Accelerator
Laboratory, Batavia, IL 60510 \\ 
$^b$Department of Physics, University of Chicago, Chicago, IL 60637\\
$^c$Department of Astronomy and Astrophysics, University of Chicago,
Chicago, IL 60637
}

\begin{abstract}

Three dark matter direct detection experiments (DAMA/LIBRA, CoGeNT, and CRESST-II) have each reported signals which are not consistent with known backgrounds, but resemble that predicted for a dark matter particle with a mass of roughly $\sim$10 GeV and an elastic scattering cross section with nucleons of $\sim$$10^{-41}$--$10^{-40}$ cm$^2$. In this article, we compare the signals of these experiments and discuss whether they can be explained by a single species of dark matter particle, without conflicting with the constraints of other experiments. We find that the spectrum of events reported by CoGeNT and CRESST-II are consistent with each other and with the constraints from CDMS-II, although some tension with xenon-based experiments remains. Similarly, the modulation signals reported by DAMA/LIBRA and CoGeNT appear to be compatible, although the corresponding amplitude of the observed modulations are a factor of at least a few higher than would be naively expected, based on the event spectra reported by CoGeNT and CRESST-II. This apparent discrepancy could potentially be resolved if tidal streams or other non-Maxwellian structures are present in the local distribution of dark matter.

\end{abstract}

\pacs{95.36.+x; FERMILAB-PUB-11-571-A}

\maketitle


\section{Introduction}

Recently, a great deal of attention has been given to a number of dark matter experiments which have reported signals that do not appear to be consistent with any known backgrounds. The longest standing of these claims is from the DAMA (and more recently the DAMA/LIBRA) collaboration, which has reported annual variation in the event rate at their lowest observed energies, in keeping with the predictions from dark matter scattering~\cite{damanew}.  Earlier this year, the CoGeNT collaboration reported that their previously observed excess of low-energy events also exhibits seasonal variation~\cite{Aalseth:2010vx,Aalseth:2011wp}, similar to the signal reported by DAMA/LIBRA. Most recently, the CRESST-II collaboration has reported an excess of events potentially attributable to dark matter~\cite{Angloher:2011uu}.

A casual comparison of these results can be confusing or even misleading. Conventionally, both experimentalists and theorists in the field of dark matter direct detection report their constraints and other results as derived using specific astrophysical assumptions and estimates of detector response. Adopting a single choice for each of these characteristics (even if that choice represents a reasonable estimate), rather than marginalizing over the possible or plausible range of those choices, can lead to regions of compatibility that are artificially small, and to constraints which are artificially stringent.

In this paper, we revisit the signals reported by the CRESST-II, CoGeNT and DAMA/LIBRA collaborations, and attempt to determine whether they can be consistently explained with dark matter. In particular, in Sec.~\ref{spectrum}, we study and directly compare the spectra of excess events reported by CoGeNT and CRESST-II, and from this conclude that (for reasonable astrophysical assumptions) a dark matter particle with a mass of roughly 10-20 GeV and an elastic scattering cross section with nucleons of approximately $(1-3)\times 10^{-41}$ cm$^2$ could account for the excess events reported by each of these collaborations. We also note that a sizable fraction of this parameter space is consistent with the constraints placed by the CDMS-II collaboration.  The constraints presented by the XENON-100 and XENON-10 collaborations, however, remain in conflict, unless either the response of liquid xenon to very low-energy nuclear recoils is lower than previously claimed, or the dark matter's couplings to protons and neutrons destructively interfere for a xenon target. In Sec.~\ref{annual}, we compare the annual modulation signals reported by the DAMA/LIBRA and CoGeNT collaborations. Again, we find good agreement between the results of these two experiments, but point out that under common assumptions (Maxwellian velocity distributions and velocity independent scattering cross sections), the amplitude of the observed modulation requires the dark matter to possess a significantly larger (by a factor of approximately 3--10) elastic scattering cross section than would be inferred from the spectra reported by CoGeNT and CRESST-II. In Sec.~\ref{streams}, we explore how this apparent discrepancy between the observed event rate and modulation amplitude could potentially be resolved by the presence of streams or other non-Maxwellian velocity structures in the local distribution of dark matter, or by dark matter with a velocity-dependent scattering cross section with nuclei. In Sec.~\ref{conclusion}, we summarize our results and draw conclusions.


\section{The Nuclear Recoil Spectrum}
\label{spectrum}

The CoGeNT and CRESST-II collaborations have each reported an excess of nuclear recoil candidate events, difficult to attribute to known backgrounds. In this section, we discuss the energy spectra of these events and the dark matter parameter space which could potentially account for these signals.

\subsection{CoGeNT's Event Spectrum}

CoGeNT is a P-type point contact germanium detector with very low levels of electronic noise, enabling sensitivity to very low-energy nuclear recoils, and thus very low mass dark matter particles. Located in Northern Minnesota's Soudan Underground Laboratory, CoGeNT observes nuclear recoil events as ionization, and has thus far reported the results of 15 months of data collection, taken between December 2009 and March 2011~\cite{Aalseth:2010vx,Aalseth:2011wp}.

\begin{figure*}[t]
\centering
{\includegraphics[angle=0.0,width=3.25in]{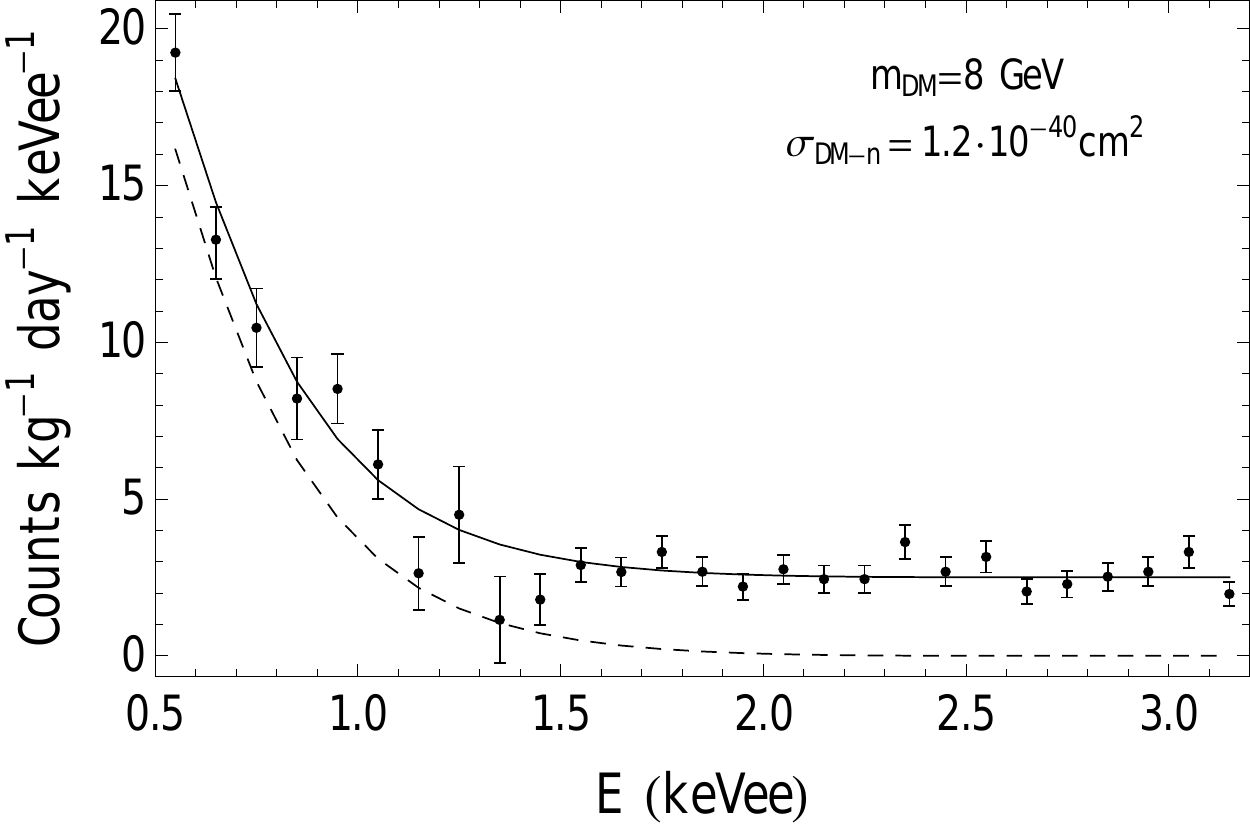}}
{\includegraphics[angle=0.0,width=3.25in]{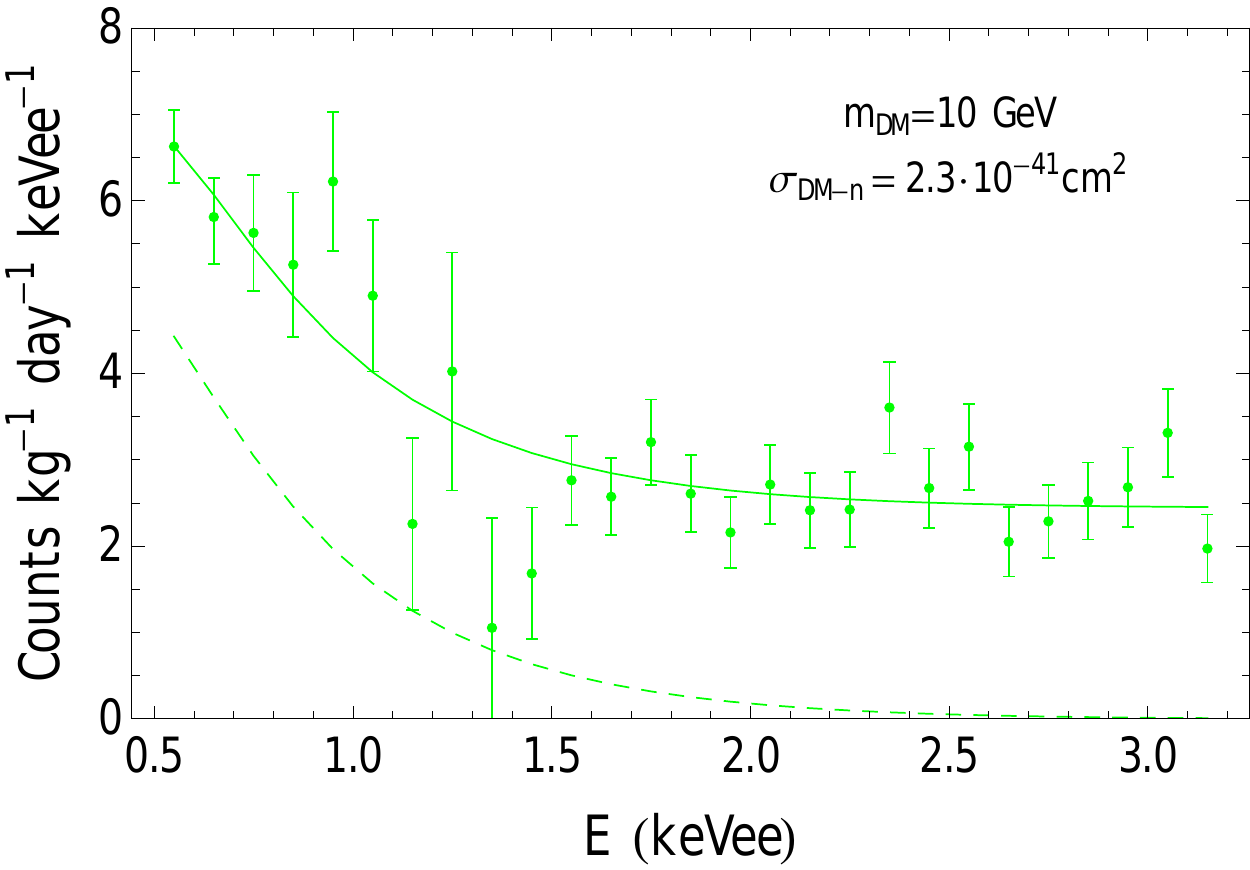}}\\
{\includegraphics[angle=0.0,width=3.25in]{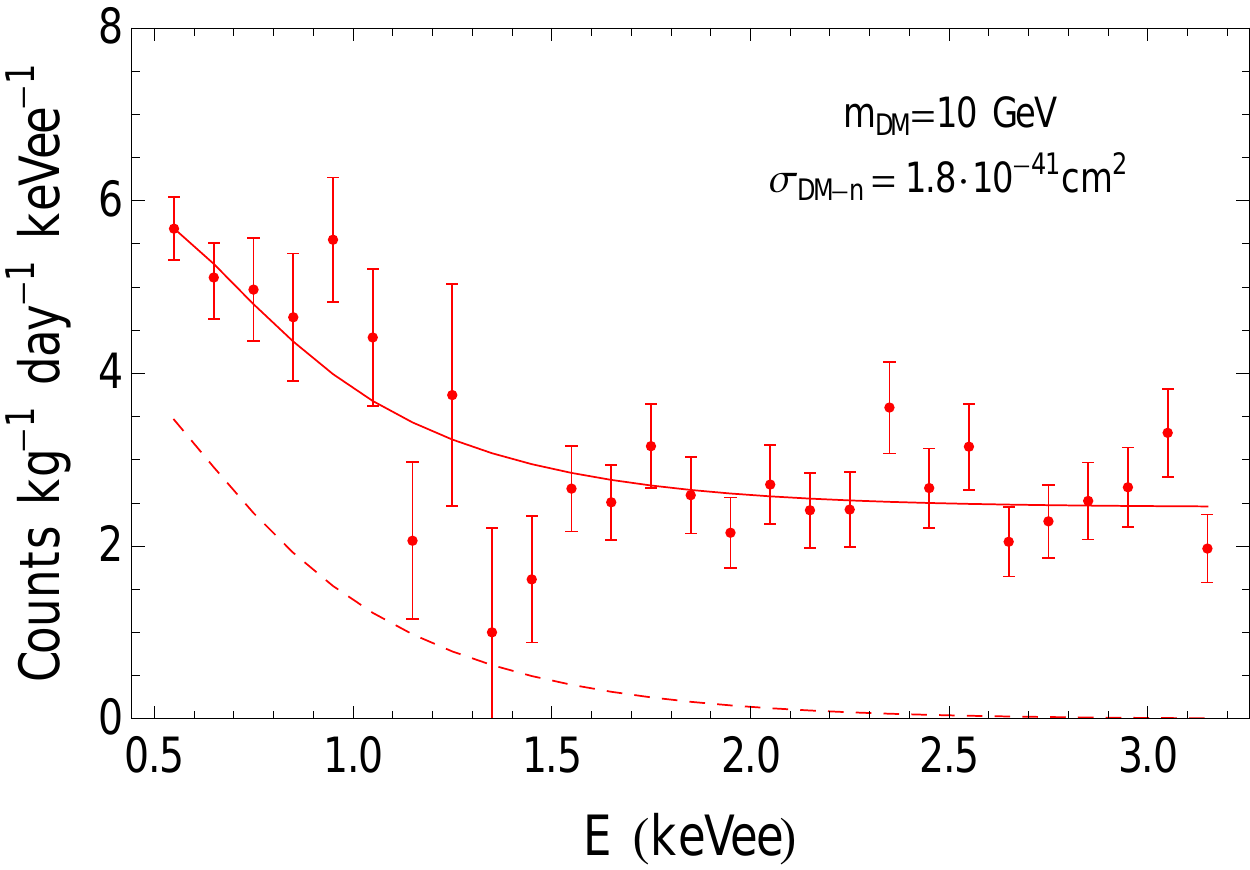}}
{\includegraphics[angle=0.0,width=3.25in]{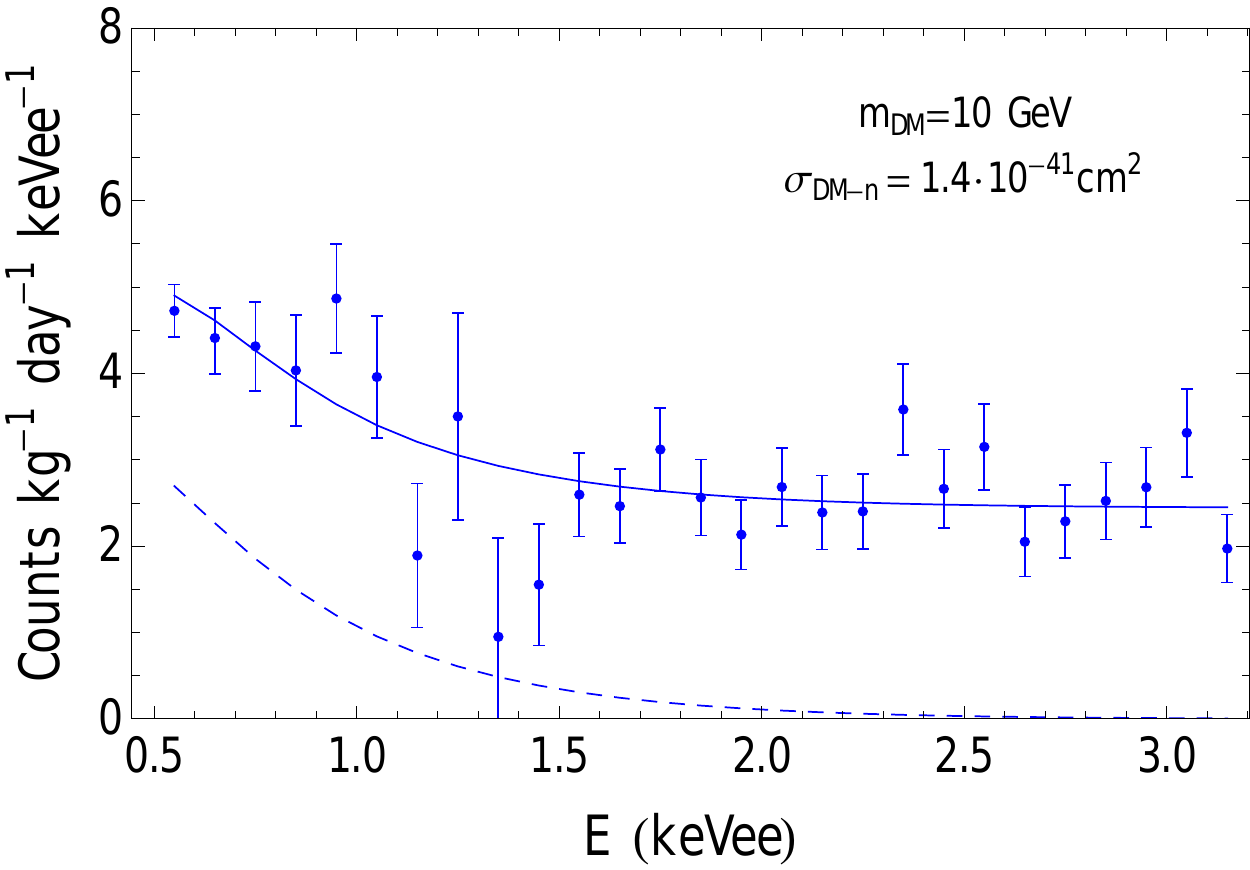}}
\caption{In the upper-left frame, we show the raw spectrum of nuclear recoil candidate events as observed by CoGeNT, as originally presented in Ref.~\cite{Hooper:2011hd}. In the other three frames, this spectrum has been corrected using three different estimates for CoGeNT's surface event correcton factor, as shown in Fig.~\ref{fig:surfEvent}.  In each frame, a spectrum of events from dark matter is shown (dashed line), along with this signal plus a flat background from Compton scattering (solid line).}
\label{fig:allSpectra}
\end{figure*}

In the upper-left frame of Fig.~\ref{fig:allSpectra}, we show the spectrum of events reported by CoGeNT as a function of ionization energy (in keV-electron equivalent, keVee), after subtracting the L-shell electron capture peaks (as described in Ref.~\cite{Hooper:2011hd}). Above about 1.5 keVee, the spectrum observed by CoGeNT is approximately flat and featureless, and is thought to be dominated by Compton scattering events. At lower energies, the observed rate climbs rapidly. While contamination from (non-rejected) surface events is expected to contribute significantly near threshold, it does not appear to be possible to account for the observed low-energy rate with this or other known backgrounds.

To assess the hypothesis that the excess events reported by CoGeNT are the product of the elastic scattering of dark matter particles, we compare CoGeNT's event spectrum to that predicted from dark matter. The spectrum (in nuclear recoil energy) of dark matter induced elastic scattering events is given by~\cite{ls}
\be
\frac{dR}{dE_R} = N_T \frac{\rho_{DM}}{m_{DM}} \int_{|\vec{v}|>v_{\rm
min}} d^3v\, vf(\vec{v},\vec{v_e}) \, \frac{d\sigma}{d E_R},
\label{rate1}
\ee
where $N_T$ is the number of target nuclei, $m_{DM}$ is the mass of
the dark matter particle, $\rho_{DM}$ is the local dark matter
density (which we take to be 0.3 GeV/cm$^3$), $\vec{v}$ is the dark matter velocity in the frame of the
Earth, $\vec{v_e}$ is the velocity of the Earth with respect 
to the Galactic halo, and $f(\vec{v},\vec{v_e})$ is the distribution
function of dark matter particle velocities, which we will take to be the standard Maxwell-Boltzmann distribution for the time being:
\be
f(\vec{v},\vec{v_e}) = \frac{1}{(\pi v_0^2)^{3/2}} {\rm
e}^{-(\vec{v}+\vec{v_e})^2/v_0^2}.
\ee
The Earth's speed relative to the Galactic halo is given by
$v_e=v_{\odot}+v_{\rm orb}{\rm cos}\,\gamma\, {\rm
cos}[\omega(t-t_0)]$ where $v_{\odot}=v_0+12\,{\rm km/s}$, 
$v_{\rm orb}=30 {\rm km/s}$, ${\rm cos}\,\gamma=0.51$, $t_0$ is the date of the peak in the annual modulation (generally anticipated to lie within several weeks of late May or early June), and $\omega=2\pi/{\rm year}$. As a default choice, we adopt the commonly used values of $v_0=220\,$m/s and a Galactic escape velocity of 544\,km/s. This function should be thought of as a reasonable, but approximate, parametrization of the dark matter's true velocity distribution. Departures from a Maxwellian velocity distribution are in fact required for consistency with observed (and simulated) halo density profiles (see, for example, Ref.~\cite{Lisanti:2010qx} and references therein). Such departures can non-negligibly impact the spectrum of dark matter induced events, and can significantly modify the degree of seasonal variation in the rate~\cite{Kuhlen:2009vh,Lisanti:2010qx}.  In Sec.~\ref{streams}, we will return to this issue and examine the extent to which non-Maxwellian structures in the velocity distribution could potentially impact the signals reported by direct detection experiments.


As the germanium isotopes which make up the CoGeNT detector contain little net spin, we consider spin-independent interactions to generate the observed events. In this case, we have
\be
\frac{d\sigma}{d E_R} = \frac{m_N}{2 v^2} \frac{\sigma_n}{\mu_n^2}
\frac{\left[f_p Z+f_n (A-Z)\right]^2}{f_n^2} F^2(q),
\label{cross1}
\ee
where $\mu_n$ is the reduced mass of the dark matter particle and
nucleon (proton or neutron), 
$\sigma_n$
 is the scattering cross section of the dark matter 
particle with 
neutrons,
$Z$ and $A$ are the atomic and mass numbers of the nucleus, and
$f_{n,p}$ are the couplings of the dark matter particle to
neutrons and protons, respectively. Unless stated otherwise, our results are calculated under the assumption that $f_p=f_n$. The nuclear form factor, $F(q)$, accounts for the finite momentum transfer in scattering events.  In our calculations, we adopt the Helm form factor with parameters as described in our previous work~\cite{kelso}. To convert from nuclear recoil energy to the measured ionization energy, we have scaled the results by the quenching factor for germanium as described in Refs.~\cite{Ge,Ge2} ($Q_{\rm Ge}=0.218$ at $E_R=$3 keV, and with the energy dependence predicted by the Lindhard theory~\cite{consistent}).

\begin{figure*}[!t]
\centering
{\includegraphics[angle=0.0,width=3.4in]{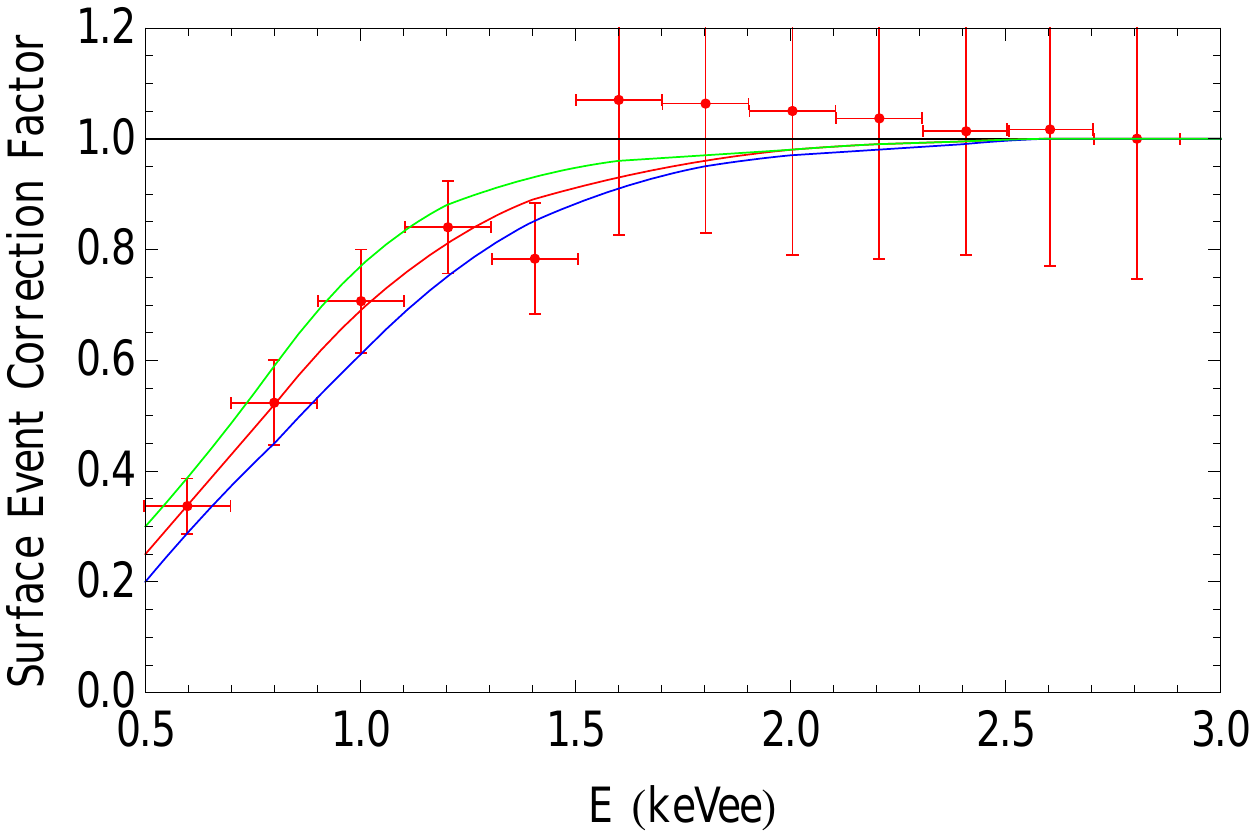}}
\caption{CoGeNT's surface event rejection correction factor (the fraction of nuclear recoil candidate events that are not surface events) as recently presented in Ref.~\cite{JuanTAUP}.  The four curves shown (including the horizontal line) correspond to the correction factors used to generate the corresponding spectra in Fig.~\ref{fig:allSpectra}.}
\label{fig:surfEvent}
\end{figure*}

Although the majority of surface events have been removed from the spectrum presented by the CoGeNT collaboration through the application of a rise-time cut, this spectrum does not take into account any inefficiencies in their surface event rejection algorithm. While it was initially estimated that the signal would suffer only a minor degree of contamination from non-rejected surface events~\cite{Aalseth:2010vx}, the CoGeNT collaboration has recently reported a somewhat higher estimate for the rate of non-rejected surface events near their energy threshold~\cite{JuanTAUP}.  This (preliminary) estimate of the fraction of the event spectrum which consists of non-surface events ({\it i.e}.~nuclear recoil candidate events) is shown in Fig.~\ref{fig:surfEvent}. Any unidentified surface events constitute an additional background that should be accounted for when attempting to identify a dark matter signal from the CoGeNT detector.  As can be observed from Fig.~\ref{fig:surfEvent}, this new measurement significantly reduces the estimate of the number of low-energy events that could potentially be attributed to dark matter. The precision of this measurement is expected to improve over time as more statistics are accumulated.

\begin{figure*}[!t]
\centering
{\includegraphics[angle=0.0,width=3in]{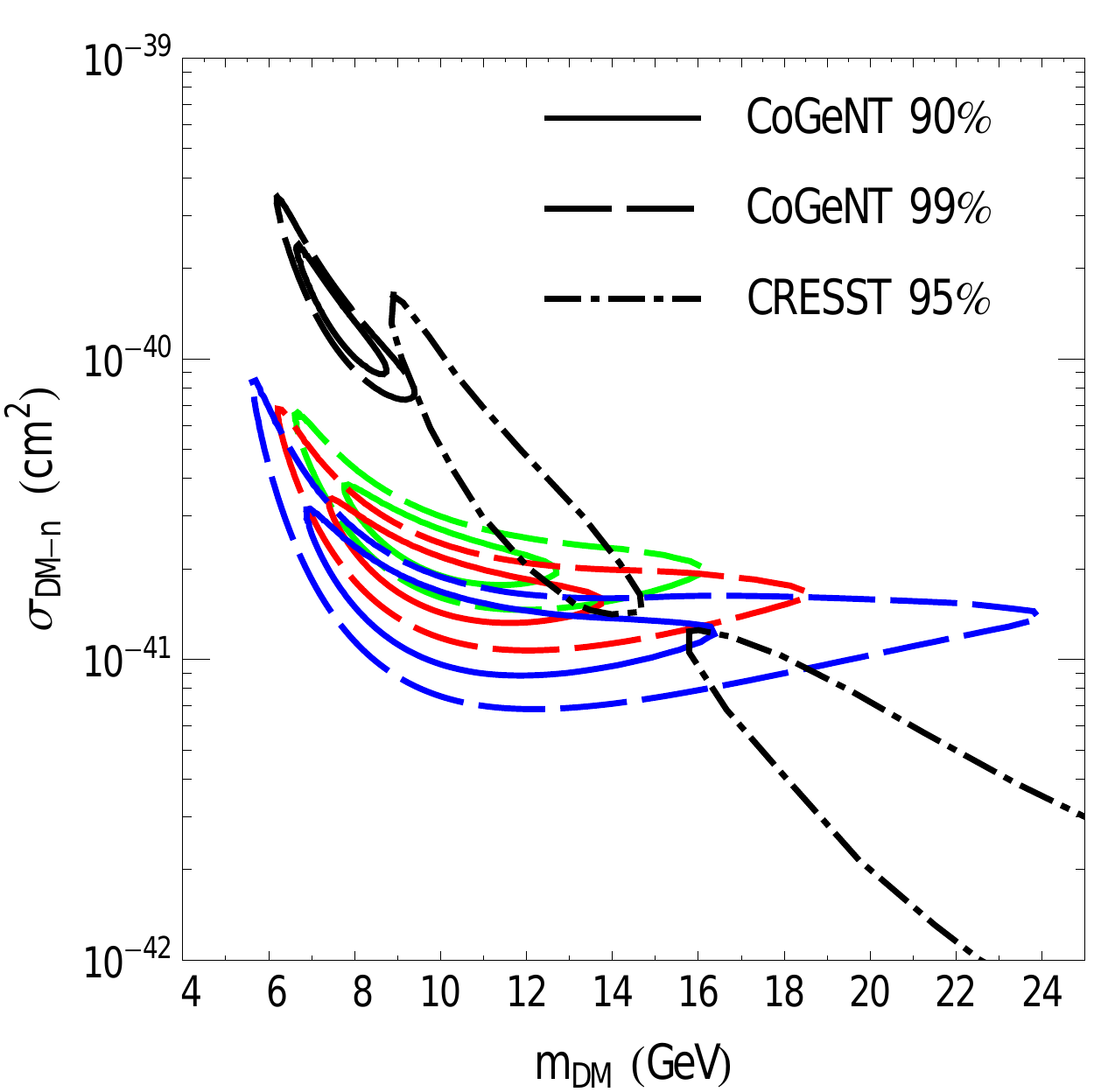}}
\caption{The 90\% (solid) and 99\% (dashed) confidence level contours for the spectrum of events observed by CoGeNT, with each color matching the corresponding correction factors shown in Fig.~\ref{fig:surfEvent}. The CRESST contours (dot-dashed) denote the 95\% confidence level regions. A dark matter particle with a mass of approximately 10-20 GeV and an elastic scattering cross section with nucleons of approximately $(1-3)\times 10^{-41}$ cm$^2$ can account for the excess events reported by each of these experiments.}
\label{fig:massCross}
\end{figure*}

In Fig.~\ref{fig:allSpectra}, we show the spectrum events at CoGeNT, for several different choices of the surface event correction factor. As mentioned previously, the upper-left frame of Fig.~\ref{fig:allSpectra} depicts the spectrum of events assuming a perfect surface event rejection efficiency (100\% of all surface events are identified as such, at all energies).  The other three frames show the remaining spectrum of events after applying the most mild (green), central (red), and most stringent (blue) correction factor, as shown in Fig.~\ref{fig:surfEvent}, to the raw spectrum.  For each of these choices of the surface event correction factor, we find that the resulting spectrum can be explained by an elastically scattering dark matter particle, although with slightly differing ranges of masses and cross sections. In each frame, we show an example of a good fit, with the dashed line denoting the signal from the dark matter alone and the solid line also including a flat background from Compton scattering events, in each case correcting for the detector efficiency, as described in Ref.~\cite{Aalseth:2010vx}.

In Fig.~\ref{fig:massCross} we plot the regions of dark matter parameter space that provide a good fit to the efficiency corrected spectra shown in Fig.~\ref{fig:allSpectra}. In each case, we have allowed the normalization of the flat background to float.  As expected, the inclusion of the non-rejected surface event background shifts the preferred region towards smaller cross sections, as there is now less dark matter signal then expected previously. The range of dark matter masses favored also shifts upward somewhat as a result of the additional surface event background.

\subsection{CRESST's Event Spectrum}

The CRESST-II collaboration makes use of eight 300 gram cryogenic CaWO$_4$ detectors, operating in Italy's Laboratori Nazionali del Gran Sasso. Due to the relatively light oxygen and calcium nuclei in their target, CRESST is quite sensitive to dark matter particles in the mass range favored by CoGeNT. CRESST observes events through both scintillation and heat (phonons), enabling them to discriminate nuclear recoil candidate events from a variety of backgrounds.

Very recently, the CRESST-II collaboration released an analysis of their first 730 kg-days of data, taken over a period between 2009 and 2011~\cite{Angloher:2011uu}. The analysis identified 67 low-energy nuclear recoil candidate events, which is at least 30\% more than can be accounted for with known backgrounds. The CRESST-II collaboration has assessed the statistical significance of this excess to be greater than 4$\sigma$.

The CRESST-II analysis identified two distinct regions of dark matter parameter space which are compatible with the observed excess (see Fig.~\ref{fig:massCross}). In the high mass region (referred to as M1), the majority of the excess events arise from dark matter recoils with tungsten nuclei. Within the low mass (M2) region, in contrast, the excess events are dominated by recoils on both oxygen and calcium. In an independent analysis based on the publicly available portions of the CRESST data, Ref.~\cite{kopp} identified a similar, but somewhat larger, region of compatible dark matter parameter space.

As can be seen in Fig.~\ref{fig:massCross}, the dark matter parameter space favored by CRESST-II is compatible with the region implied by CoGeNT's spectrum, after correcting for surface event contamination. In particular, a dark matter particle with a mass of roughly 10-20 GeV and an elastic scattering cross section with nucleons of $(1-3)\times 10^{-41}$ cm$^2$ could account for the excess events reported by both collaborations. 



\begin{figure*}[!t]
\centering
{\includegraphics[angle=0.0,width=3.4in]{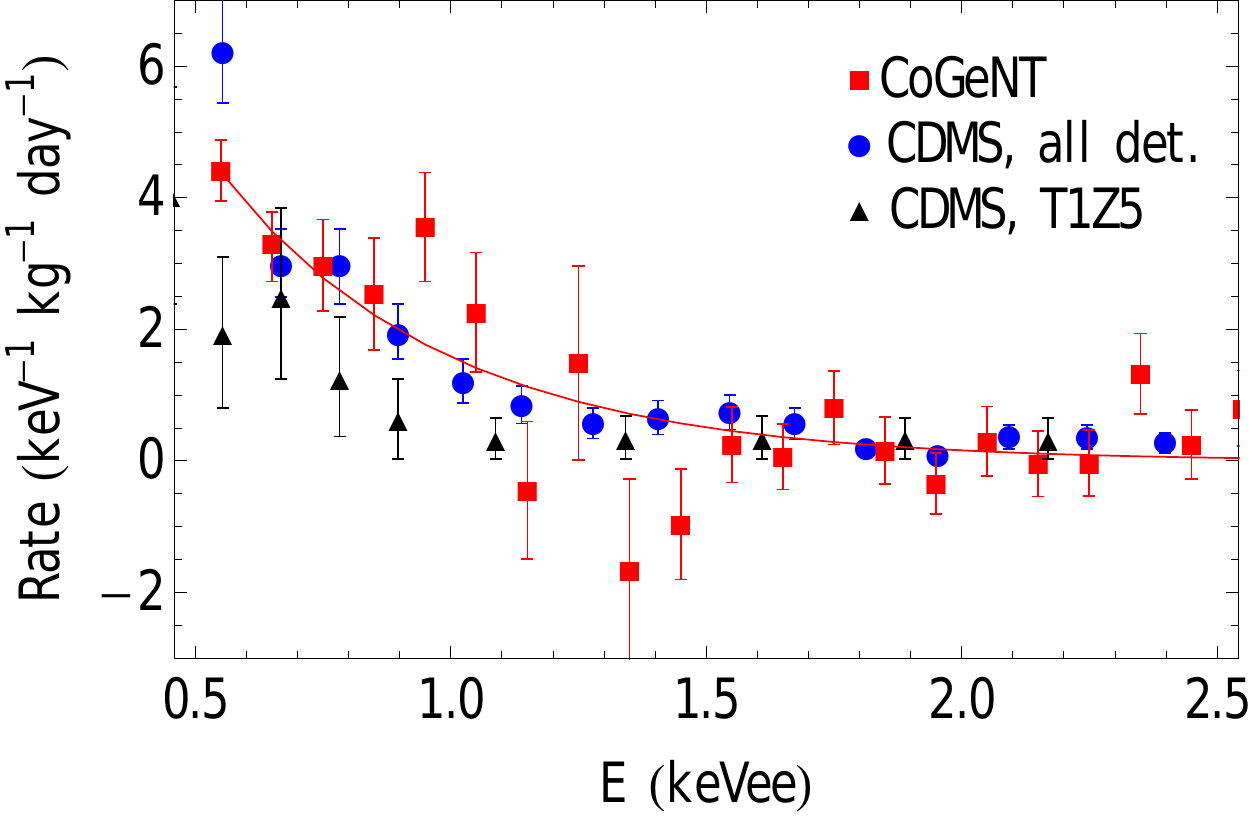}}
\caption{A comparison of CoGeNT's spectrum (using the central estimate for the surface event correction factor, as shown in Fig.~\ref{fig:surfEvent}) with that reported by the CDMS-II collaboration~\cite{Ahmed:2010wy}. We have subtracted the flat (Compton scattering) component from CoGeNT's spectrum, and corrected for CoGeNT's efficiency. The solid curve represents the prediction for a 10 GeV dark matter particle with an elastic scattering cross section of $\sigma_{n}=1.8\times 10^{-41}$ cm$^2$. The spectrum observed by the combination of all of CDMS's detectors is in good agreement with that observed by CoGeNT, although the spectrum from the single detector, T1Z5, is slightly lower than CoGeNT's below 1.2 keVee.}
\label{fig:cdms}
\end{figure*}

\subsection{Constraints From Other Experiments}
\label{exclusion}

A number of direct detection experiments have produced constraints which are relevant to the interpretation of the events reported by CoGeNT and CRESST-II. In particular, the impact of the constraints presented by the CDMS-II, XENON-100, and XENON-10 collaborations are significant for the regions of low-mass dark matter parameter space favored by CoGeNT and CRESST-II.

The CDMS-II collaboration has presented the results of two analyses searching for dark matter particles in the mass range collectively favored by CoGeNT and CRESST~\cite{Ahmed:2010wy,Akerib:2010pv}. Before taking into account the updated estimates of CoGeNT's surface event rejection efficiency, these constraints appeared to be in conflict with a dark matter interpretation of CoGeNT's excess (see, however, Ref.~\cite{Collar:2011kf}). As both CDMS and CoGeNT make use of germanium detectors, and thus are sensitive to similar systematic factors such as quenching factors for low-energy nuclear recoils, it was generally considered difficult to reconcile CDMS's constraints with a dark matter interpretation of CoGeNT. In light of the CoGeNT collaboration recent estimate for their surface event rejection efficiency, however, this apparent conflict seems to be largely resolved. In Fig.~\ref{fig:cdms}, we compare the spectrum at CoGeNT (after subtracting the flat, Compton scattering, component, and applying the central estimate for the surface event correction factor) to that reported by the low-threshold analysis of CDMS-II. While the spectrum below 1.2 keVee from CDMS's T1Z5 detector is slightly lower than that observed by CoGeNT, the all-detectors spectrum reported by CDMS is in good agreement with CoGeNT's.

The XENON-100~\cite{xenon100} and XENON-10~\cite{xenon10} collaborations have also each reported rather strong constraints on the parameter space of low-mass dark matter particles. As presented, these constraints appear to largely rule out the dark matter parameter space collectively favored by CoGeNT and CRESST. There are a number of ways, however, in which these constraints could be significantly weaker than they might appear. Firstly, any uncertainties in the response of liquid xenon to very low-energy nuclear recoils (as encapsulated in the functions $L_{\rm eff}$ and/or $Q_y$) could significantly impact the corresponding constraints for dark matter particles with a mass in the range of interest. The constraints from the XENON-100 collaboration were derived using measurements of the scintillation efficiency, $L_{\rm eff}$, as described in Refs.~\cite{plante}, which have been criticized in Ref.~\cite{juan2} (see also Ref.~\cite{Collar:2010ht}). Even modest changes to these values at the lowest measured energies ($\sim$3-4 keV) can lead to much weaker constraints on light dark matter particles. It has also been argued that the relatively large (9.3 eV) band-gap of xenon is expected to lead to a suppression of the response to nuclear recoils in the energy range of interest (see Ref.~\cite{Collar:2010ht} and references therein). Many of these issues also apply to constraints on light dark matter making use of only the ionization signal in liquid xenon detectors~\cite{xenon10}.

Alternatively, the constraints from XENON-100 and XENON-10 could be modified if dark matter particles do not have identical couplings to protons and neutrons~\cite{zurek,feng}. In particular, for a ratio of couplings given by $f_n/f_p \approx -0.7$, the constraint from xenon-based experiments is weakened by a factor of $\sim$20 relative to that found in the $f_n=f_p$ case~\cite{feng}. For this ratio of couplings, the cross section favored by CRESST-II would also be moved down by a factor of $\sim$7 relative to that observed by CoGeNT. Alternatively, a ratio of $f_n/f_p \approx -0.6$ would reduce the strength of the XENON-100 and XENON-10 constraints by a factor of 3-4, while also lowering the CRESST-II region (relative to that of CoGeNT) by a similar factor.

Lastly, we note that a constraint has also been placed by making use of the CRESST commissioning run data~\cite{Brown:2011dp}. These results appear to be in mild tension with the upper range (in cross section) of the parameter space reported to be favored by the analysis of the CRESST-II collaboration.

\section{Annual Modulation}
\label{annual}


If a population of events observed in a detector are in fact the result of elastically scattering dark matter particles, then we should expect the Earth's motion around the Sun to induce a degree of seasonal variation in the rate of those events. For most commonly assumed velocity distributions of dark matter particles, the rate is predicted to follow a roughly sinusoidal behavior, with a peak that occurs within several weeks of late May or early June~\cite{modulation}.

The CoGeNT and DAMA/LIBRA collaborations have each reported the observation of annual modulation of their event rates. In this section, we characterize and compare the modulation signals reported by these collaborations, and discuss whether these signals could be the result of dark matter. 

\subsection{DAMA/LIBRA's Modulation}

The DAMA/LIBRA experiment makes use of a large mass detector (242.5 kg in its current form) consisting of high purity NaI(Tl) crystals, located at Gran Sasso. DAMA/LIBRA observes nuclear recoil events as scintillation, and is designed to search for time variations in their event rate, rather than to identify individual dark matter candidate events. 

Based on the data collected over a period of 13 annual cycles, the DAMA/LIBRA collaboration reports evidence of an annual modulation with a statistical significance of 8.9$\sigma$. The variation of their rate is consistent with a sinusoid peaking at May 16$\pm$7 days at energies between 2 and 4 keV, May 22$\pm$7 days between 2 and 5 keV, and May 26$\pm$7 days between 2 and 6 keV, consistent with that predicted for dark matter with a roughly Maxwellian velocity distribution.

While DAMA/LIBRA's strategy of looking for an annual modulation in their rate can be successfully used to separate a dark matter signal from many possible backgrounds, one might worry about sources of background which could also exhibit seasonal variation. For example, the underground muon flux is known to modulate as a result of temperature variations in the stratosphere (although with a later phase and lower rate than is observed by DAMA/LIBRA~\cite{Bernabei:2009du}). Observed variations in the radon-induced background rate are also out-of-phase with the signal reported by DAMA/LIBRA. To date, no background has been identified with a phase, spectrum and rate compatible with DAMA/LIBRA's signal.

\begin{figure*}[!t]
\centering
{\includegraphics[angle=0.0,width=3.0in]{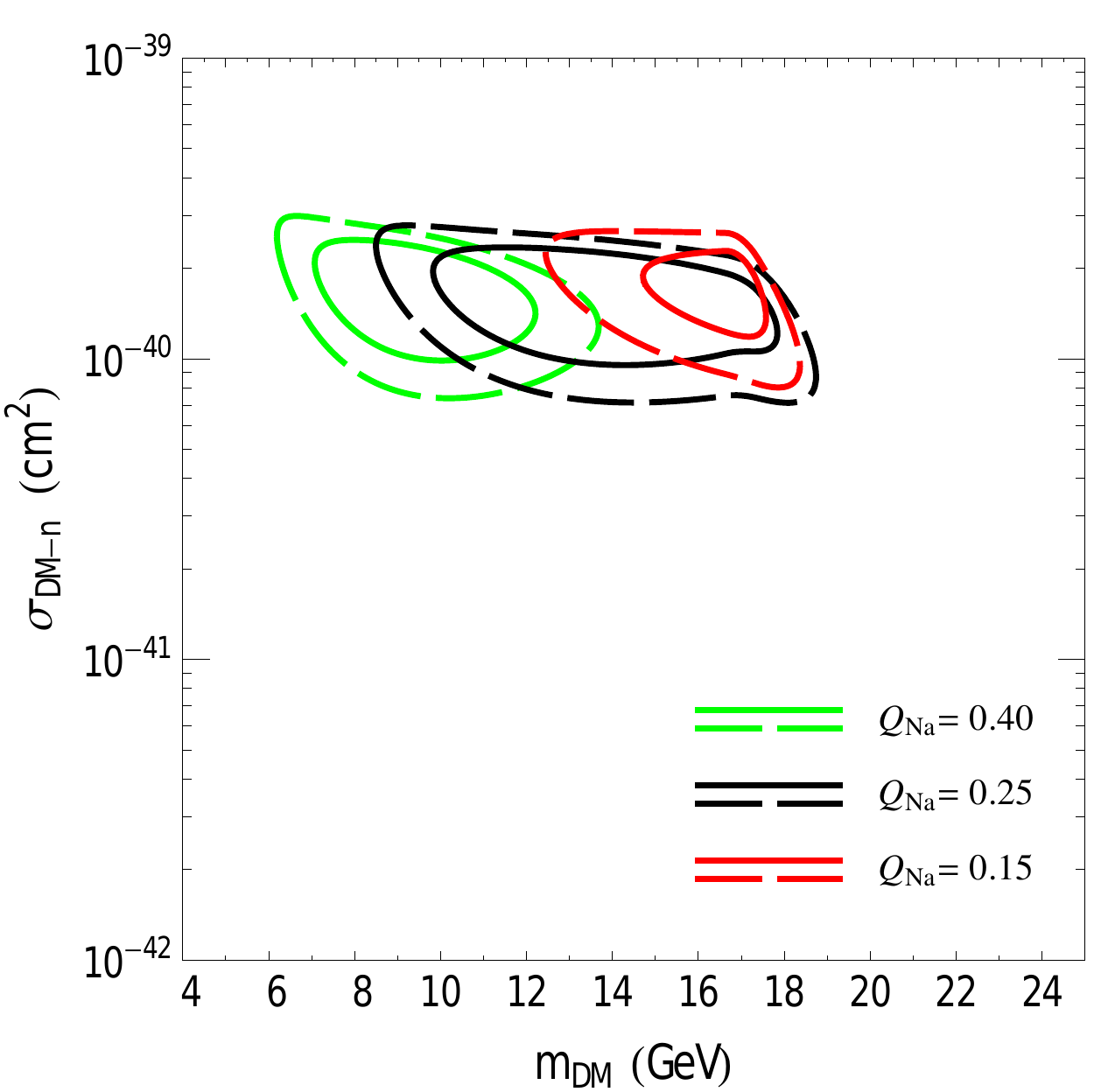}}
\caption{The 90\% (solid) and 99\% (dashed) confidence level contours for the spectrum of the amplitude of the annual modulation observed by DAMA/LIBRA, assuming a simple Maxwellian velocity distribution with $v_0=220\,$km/s and $v_{\rm esc}=544\,$km/s. Contours are shown for three choices of the low-energy sodium quenching factor, $Q_{\rm Na}$.}
\label{fig:dama}
\end{figure*}

The regions of dark matter parameter space in which the DAMA/LIBRA modulation can be accounted for depends strongly on the highly uncertain low-energy sodium quenching factor (the fraction of recoil energy of an elastic scattering event which is manifest as scintillation). In their analysis, the DAMA/LIBRA collaboration has adopted a canonical value of $Q_{\rm Na}=0.3 \pm 0.01$ for this quantity, which they report to be the measured value averaged over the recoil energy range of 6.5 to 97 keV~\cite{damaquenching}. Other groups have reported similar values:
$Q_{\rm Na} = 0.25 \pm 0.03$ (over 20-80 keV), $0.275 \pm 0.018$
(over 4-252 keV), and $0.4 \pm 0.2$ (over 5-100
keV)~\cite{Fushimi:1993nq}. As the sodium quenching factor is
generally anticipated to vary as a function of energy, it is very
plausible that over the range of recoil energies relevant for light
(5-20 GeV) dark matter particles (approximately 5 to 30 keV) the quenching
factor could be quite different from the average values reported from
these measurements. For recoil energies below approximately
20 keV, Ref.~\cite{Tovey:1998ex} reports a measurement of $Q_{\rm Na}
= 0.33 \pm 0.15$, whereas Ref.~\cite{spooner} reports a somewhat
smaller value of $Q_{\rm Na}=0.252 \pm 0.064$ near 10 keV. The results of a very recent and preliminary measurement favor values of $Q_{\rm Na} \approx 0.15-0.2$ at similarly low energies~\cite{JuanTAUP,JuanPC}. At this time, we choose to keep an open mind regarding the relevant low-energy quenching factor for sodium, and will consider a range of values between $Q_{\rm Na}\sim 0.15$ and 0.40. Based on theoretical considerations~\cite{nochanneling,Savage:2010tg} and recently experimental evidence~\cite{JuanTAUP}, we do not consider the possibility that channeling plays an important role at DAMA/LIBRA.

For a quenching factor of $Q_{\rm Na}\approx 0.25$ (0.15, 0.4), elastically scattering dark matter with a mass in the range of approximately 8-19 GeV (12-19 GeV, 6-14 GeV) can accommodate the spectrum of the modulation amplitude reported by DAMA/LIBRA (see Fig.~\ref{fig:dama}), assuming a Maxwellian velocity distribution with typical parameters (for earlier fits of DAMA data to light dark matter particles, see Refs.~\cite{Bottino:2003cz,Gondolo:2005hh}). The allowed regions do not extend to masses above about 18-20 GeV, where scattering with iodine nuclei begins to dominate. Under these same assumptions, an elastic scattering cross section of $\sigma_{n}\approx (0.7-3) \times 10^{-40}$ cm$^2$ is required to produce the observed magnitude of DAMA/LIBRA's modulation, which is significantly larger than the cross section implied by the spectra reported by CoGeNT and CRESST. Departures from a Maxwellian velocity distribution, however, could strongly impact (and potentially enhance) the observed modulation amplitude. We will return to this issue in Sec.~\ref{streams}.

\begin{figure*}[!t]
\centering
{\includegraphics[angle=0.0,width=2.4in]{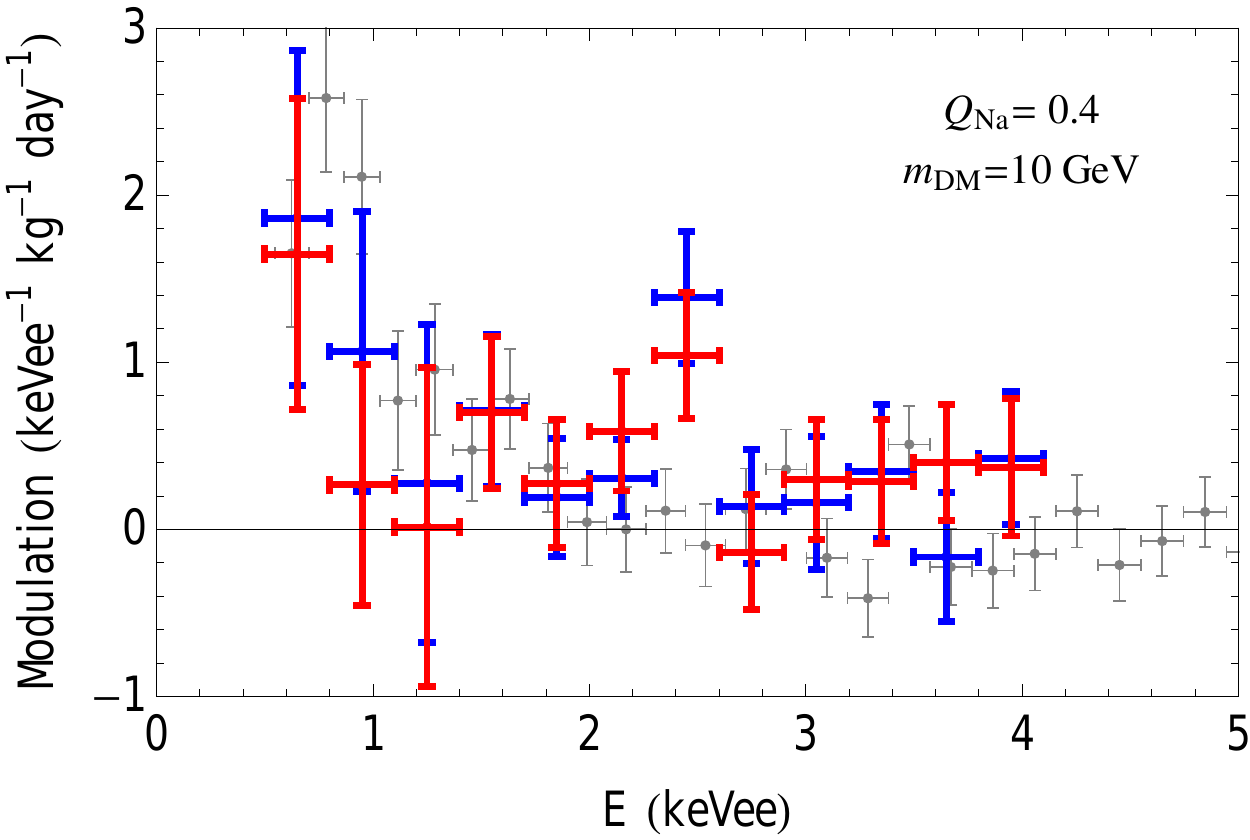}}
{\includegraphics[angle=0.0,width=2.25in]{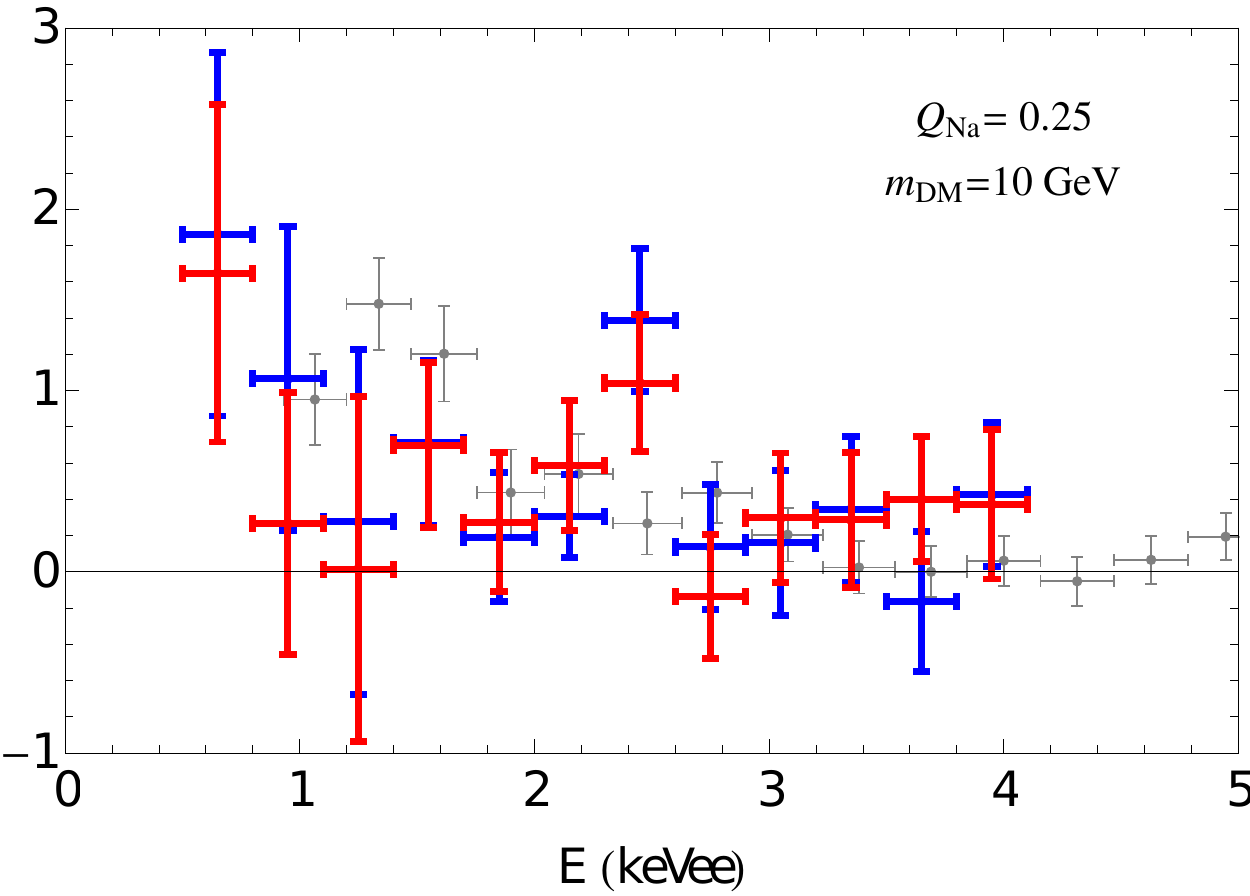}}
{\includegraphics[angle=0.0,width=2.25in]{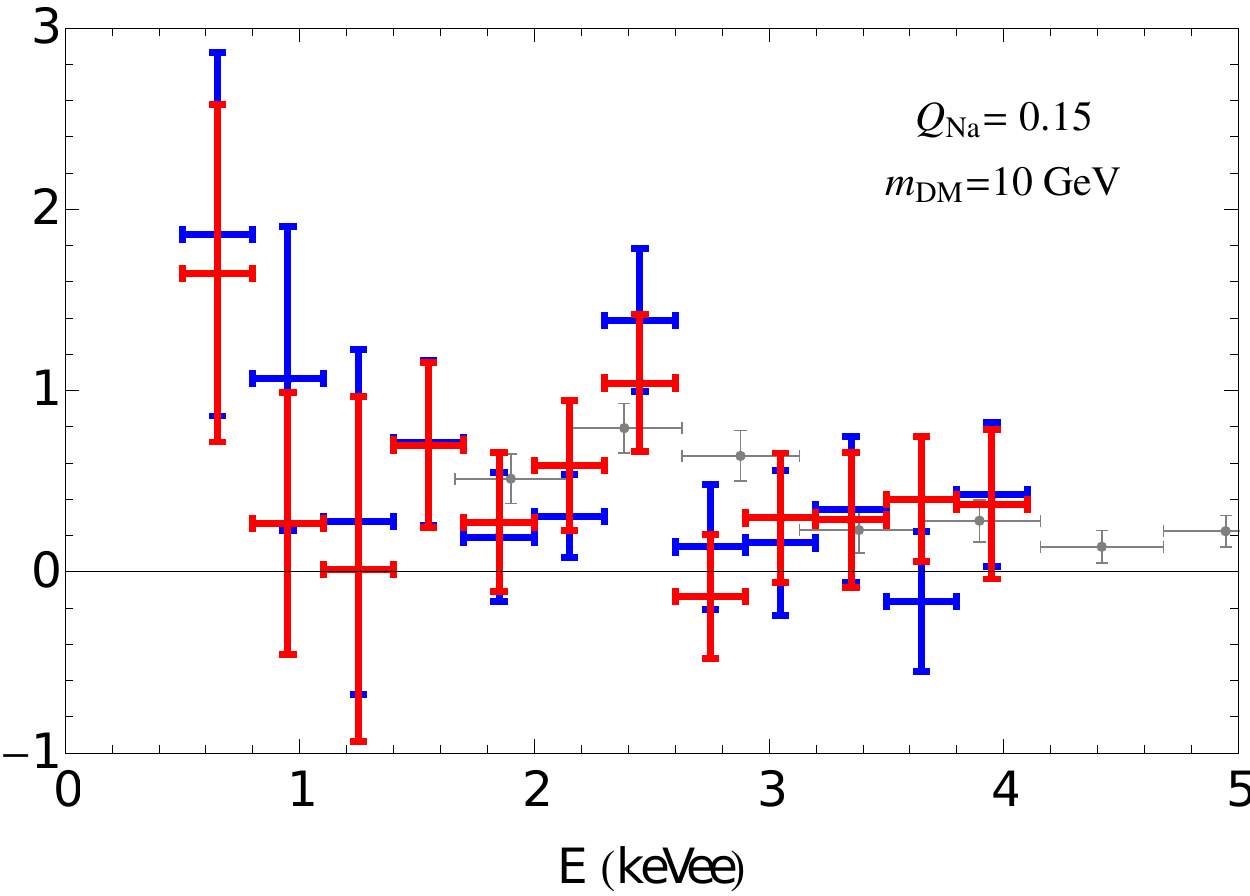}}\\
{\includegraphics[angle=0.0,width=2.4in]{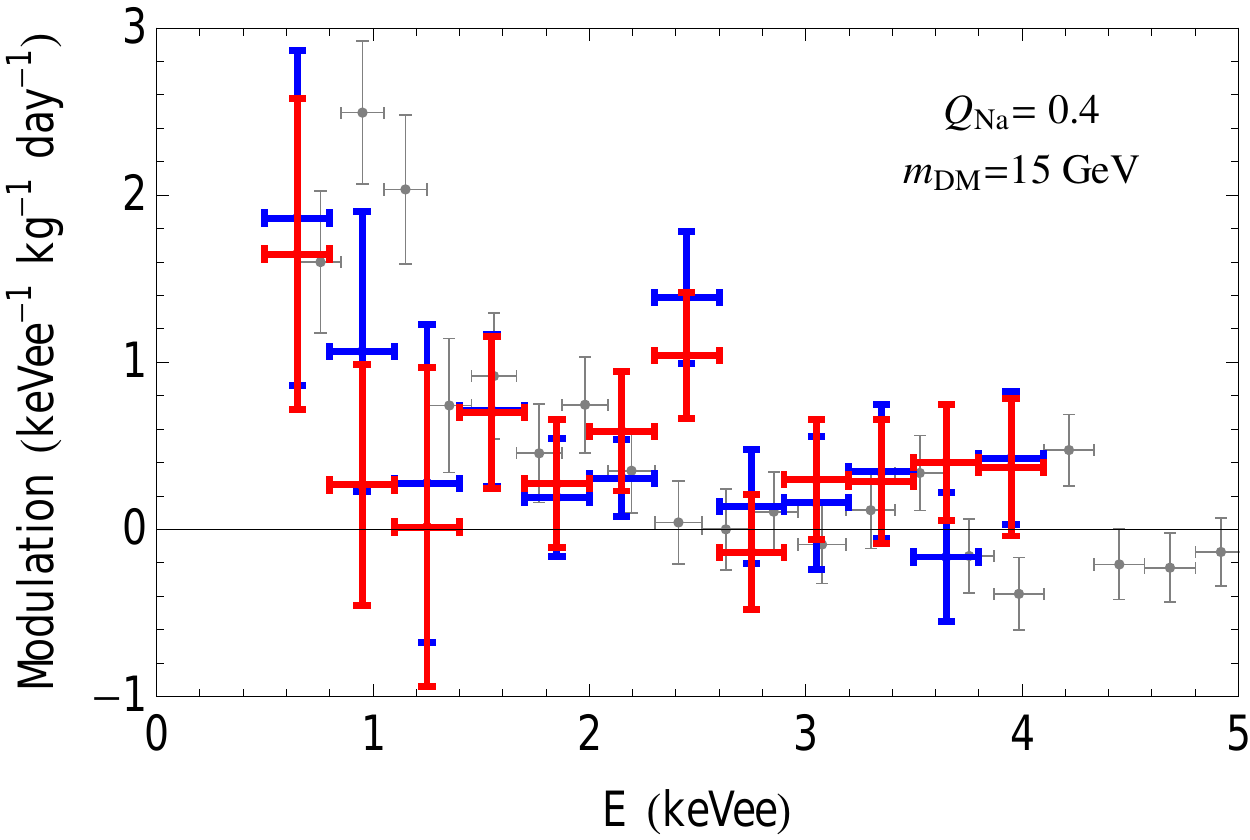}}
{\includegraphics[angle=0.0,width=2.25in]{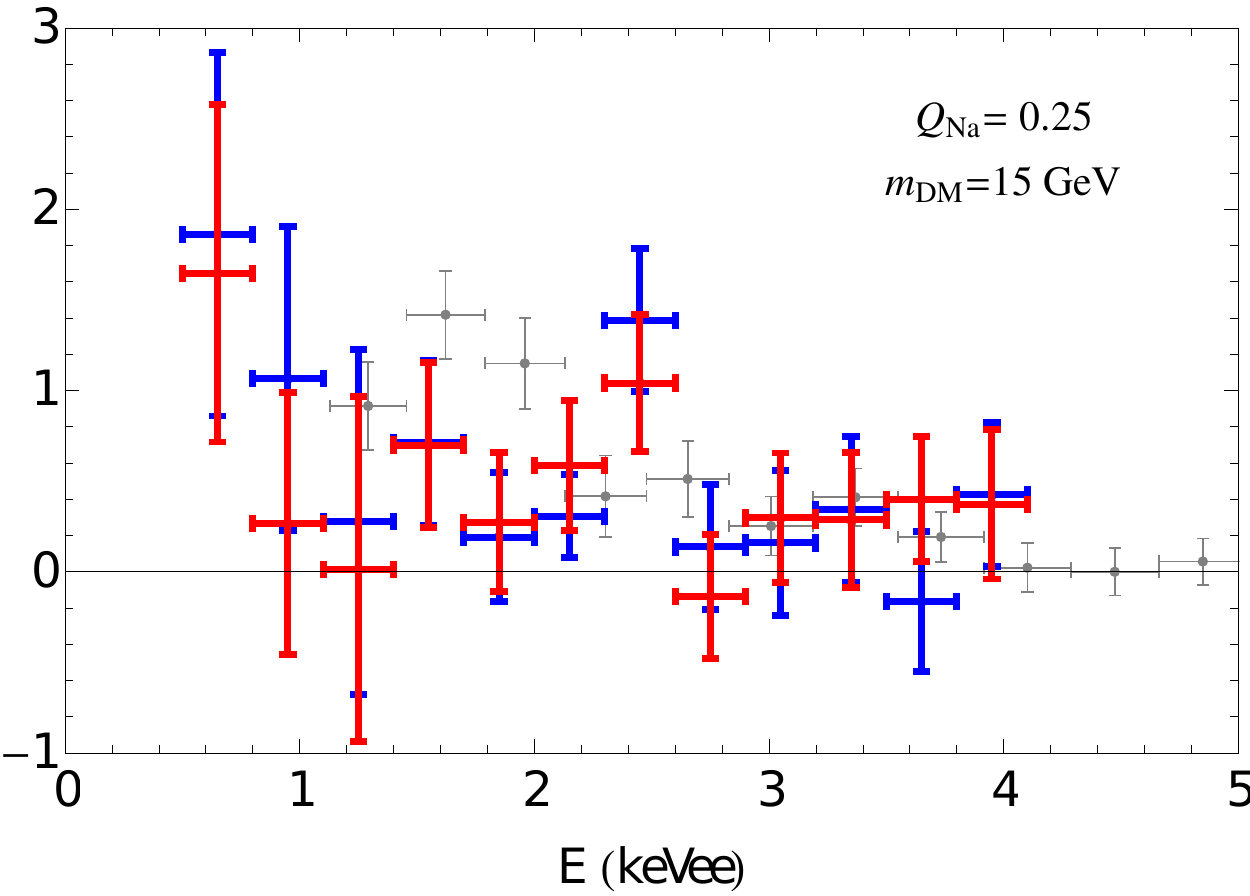}}
{\includegraphics[angle=0.0,width=2.25in]{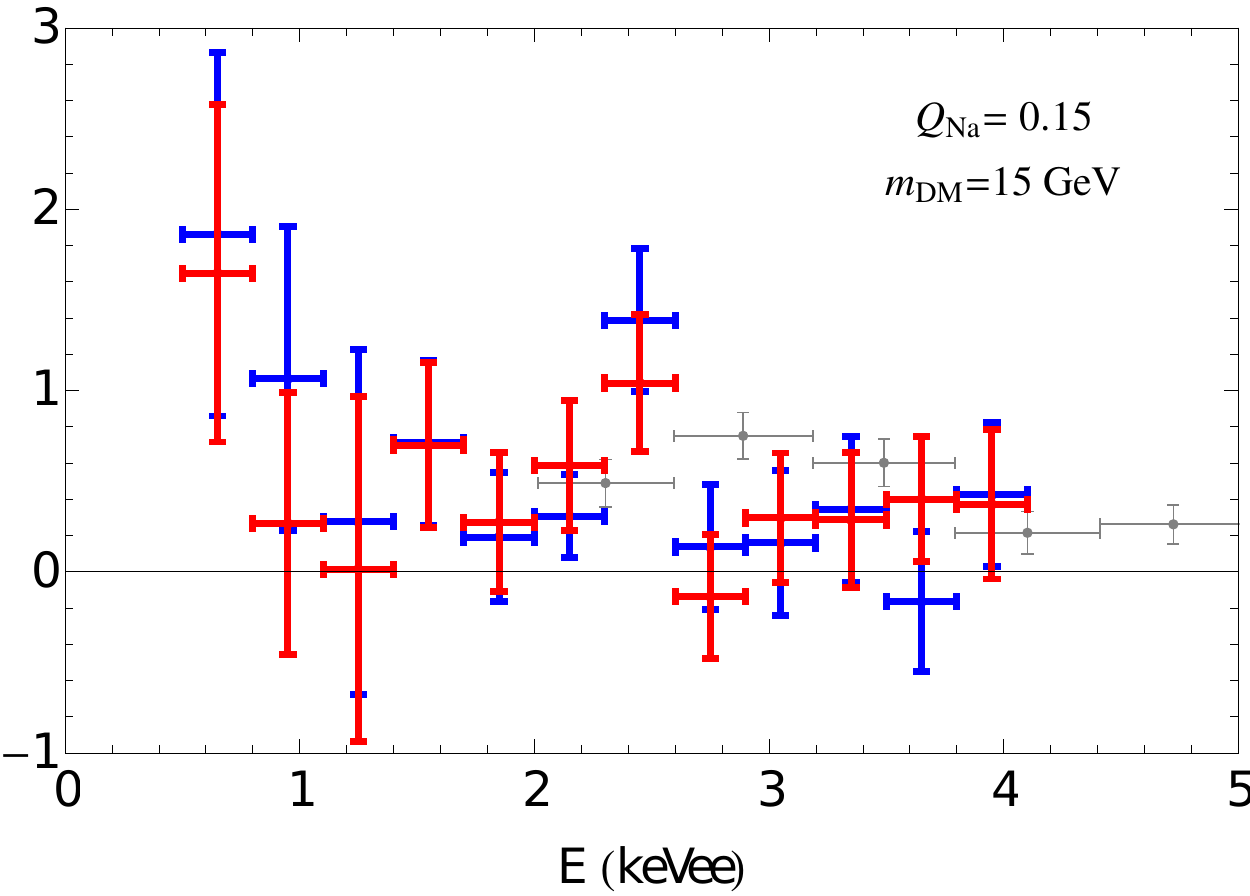}}
\caption{A comparison between the modulation amplitude spectrum observe by DAMA/LIBRA and CoGeNT, independent of the dark matter's velocity distribution, following the approach of Ref.~\cite{Fox:2010bz}. The comparison is done for a dark matter masses of 10 and 15 GeV, and for three choices of the low-energy sodium quenching factor. The blue (red) error bars denote the CoGeNT modulation amplitude assuming a phase that peaks on April 18th (May 26th). The grey error bars denote the DAMA/LIBRA modulation spectrum. In normalizing the results, we have assumed the dark matter's elastic scattering cross section to scale with the square of the target's atomic number, $A^2$.}
\label{fig:damaAtCogent}
\end{figure*}

\subsection{CoGeNT's Modulation}

The CoGeNT collaboration has also reported evidence of an annual modulation in their event rate, although with a modest statistical significance of 2.8$\sigma$. Despite the lower statistical significance of this signal, it is interesting to compare the features of CoGeNT's time variation with that observed by DAMA/LIBRA. The peak of CoGeNT's phase is May 18$\pm$16 days, which is slightly earlier (at the 1.6$\sigma$ level) than that favored by DAMA/LIBRA. If the modulation signals reported by DAMA/LIBRA and CoGeNT both arise from dark matter, a common phase that peaks in early May seems most likely~\cite{Hooper:2011hd}, in concordance with expectations for dark matter based on results from numerical simulations~\cite{Kuhlen:2009vh}. 

In comparing the spectra of the modulation amplitudes reported by DAMA/LIBRA and CoGeNT, it is possible to remove the dependence on the dark matter's velocity distribution, following the approach of Ref.~\cite{Fox:2010bz} (see also Ref.~\cite{Fox:2010bu}). In Fig.~\ref{fig:damaAtCogent}, we present such a comparison. Although this comparison does not depend on the velocity distribution of the dark matter particles, it does rely on assumptions pertaining to the mass of the dark matter particle, on the ratio of the elastic scattering cross sections with germanium and sodium, and on the relevant quenching factors. Based on the shape of the CoGeNT and CRESST-II event spectra, we choose here to consider masses of 10 and 15 GeV, and assume a cross section which scales with $A^2$ of the target nucleus, as predicted for generic spin-independent scattering. In each of the three frames, we show the results for a different value of the low-energy sodium quenching factor. The spectrum of CoGeNT's modulation amplitude was determined using the (publically available) 15 month CoGeNT data set~\cite{Aalseth:2011wp}, as described in Ref.~\cite{Hooper:2011hd}.




From Fig.~\ref{fig:damaAtCogent}, it is immediately evident that the spectrum and overall normalization of the modulation amplitudes reported by CoGeNT and DAMA/LIBRA are quite similar. In fact, if the modulation reported by DAMA/LIBRA is the product of spin-independent elastic scattering with dark matter, then one should expect CoGeNT to observe a modulation with broad features very much like that they report, and vice versa. The details of this comparison, however, depend significantly on the value of the low-energy sodium quenching factor that is adopted. For a larger value ($Q_{\rm Na}\approx 0.3-0.4$) the steadily increasing modulation amplitude at low energies is seen by both experiments, while CoGeNT's high modulation bin at $\sim$2.5 keVee is not confirmed by DAMA/LIBRA. As this feature is apparent only in one bin with a sizable error bar, we consider it possible that this bin represents a statistical fluctuation which may disappear with further data from CoGeNT. Alternatively, if a lower sodium quenching factor is adopted ($Q_{\rm Na}\approx 0.15-0.2$, as favored by Ref.~\cite{JuanTAUP,JuanPC}), the modulation reported by DAMA/LIBRA can overlap with CoGeNT's 2.5 keVee bin, while CoGeNT's lower energy modulation falls below the energy threshold of DAMA/LIBRA. In this case, one could consider the possibility that this narrow feature results from a velocity of stream of dark matter present in the local halo.

\section{Why Is The Observed Modulation Amplitude So Large?}

In the previous sections, we found that the event spectra observed by CoGeNT and CRESST-II are compatible with arising from the same dark matter particle. Similarly, the modulation amplitudes reported by DAMA/LIBRA and CoGeNT appear to be mutually consistent. Under the standard assumptions of a Maxwellian velocity distribution and velocity-independent scattering cross sections, however, the spectrum and rate of events reported by CoGeNT and CRESST-II would lead one to expect a signficantly smaller (by a factor of 3--10) modulation amplitude than is observed by DAMA/LIBRA and CoGeNT. In this section, we discuss how departures from these assumptions could explain why DAMA/LIBRA and CoGeNT have observed more modulation than would be naively predicted.


\subsection{Streams and Other Non-Maxwellian Velocity Distributions}
\label{streams}

\begin{figure*}[!t]
\centering
{\includegraphics[angle=0.0,width=2.4in]{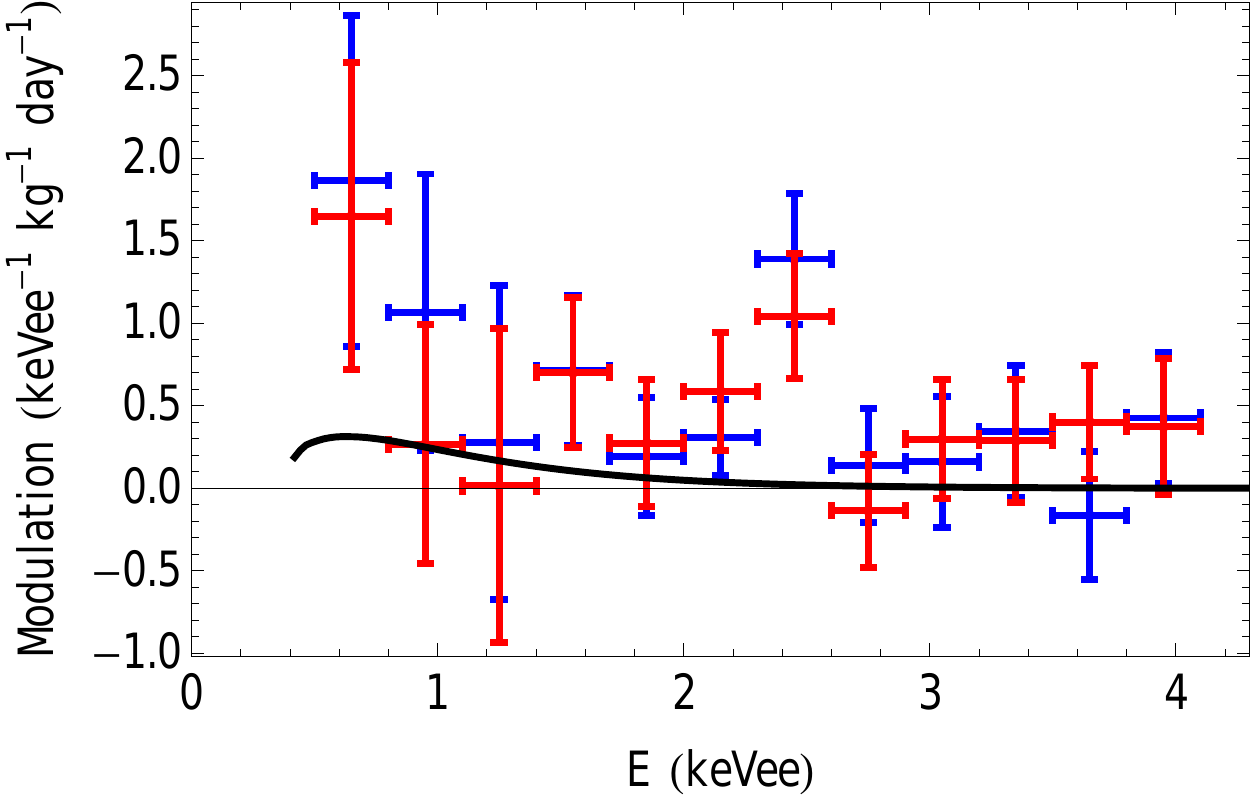}}
{\includegraphics[angle=0.0,width=2.25in]{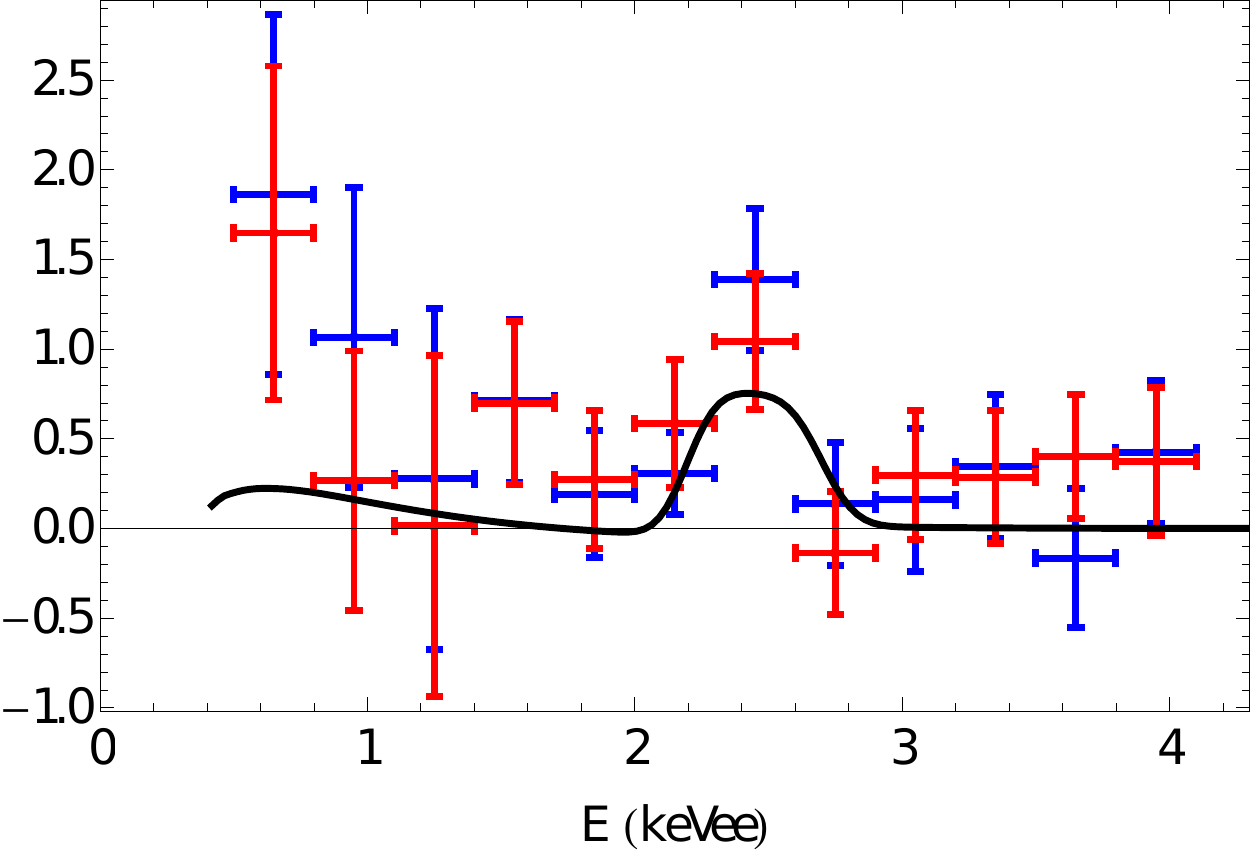}}
{\includegraphics[angle=0.0,width=2.25in]{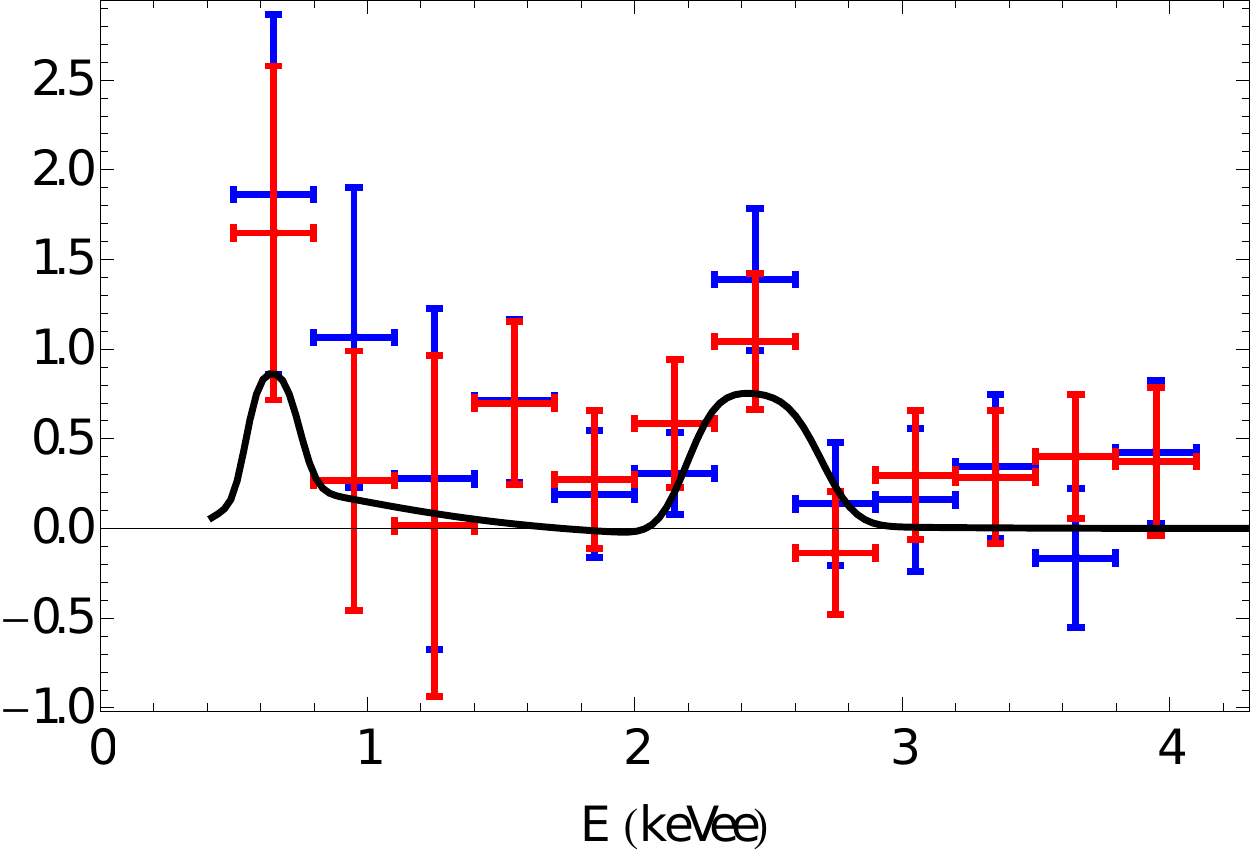}}
\caption{The impact of tidal streams on the modulation spectrum measured at CoGeNT. In the left frame, the result of a simple Maxwellian distribution is shown, for a dark matter mass of 10 GeV and an elastic scattering cross section of $\sigma_n=3 \times 10^{-41}$ cm$^2$. In the center frame, we add a tidal stream of dark matter, with a velocity of 475 km/s, a local density of 0.072 GeV/cm$^3$ (24\% of the density of the smooth halo), and a dispersion of 15 km/s. The characteristics of this stream were chosen to explain CoGeNT's rather high error bar at 2.5 keVee (and, for the appropriate choice of the sodium quenching factor, the peak in the DAMA/LIBRA spectrum as well; see the upper right frame of Fig.~\ref{fig:damaAtCogent}). In the right frame, we add a second stream, with a velocity of 165 km/s, a local density of 0.045 GeV/cm$^3$ (15\% of the density of the smooth halo), and a dispersion of 15 km/s.}
\label{fig:stream}
\end{figure*}

Numerical simulations of the formation and evolution of Milky Way-like dark matter halos have become increasingly sophisticated in recent years. These simulations find that although simple halo models with Maxwellian distributed velocities are likely to represent a reasonable zeroth order description of the distribution of dark matter in our galaxy, significant departures from such models are to be expected~\cite{nonmaxwell}.

When considering relatively light dark matter particles, as we are in this paper, the behavior of the velocity distribution near the escape velocity of the galaxy is of particular importance.  In order to produce a measurable nuclear recoil, a low-mass dark matter particles requires greater speeds than would be necessary for a heavier particle. As a result, a detector may only be sensitive to the high velocity tails of the dark matter distribution, where departures from Maxwellian behavior are expected to be most significant~\cite{Lisanti:2010qx}.

Small-scale structure of the Milky Way's halo can also play an important role in interpreting signals from direct detection experiments. The dark matter halo of our galaxy formed through a sequence of mergers of many smaller halos; a process known as hierarchial structure formation. High resolution simulations have found that many smaller halos survive this process and remain intact today, residing as substructures within larger halos~\cite{simulations}. Furthermore, many of these subhalos have a great deal of their outer mass stripped, resulting in the formation of cold tidal streams. The presence of such streams in or around the Solar System would introduce departures from the Maxwellian velocity distributions. While such streams could potentially effect the spectrum of dark matter-induced events that are observed in direct detection experiments~\cite{nonmaxwelldirect}, these effects are often far more pronounced in the modulation signals of such experiments~\cite{Kuhlen:2009vh}.  The presence of such streams can significantly enhance a modulation signal, as well as shift the phase of the modulation relative to that predicted in more simple halo models~\cite{Kuhlen:2009vh,ls}.



In Fig.~\ref{fig:stream}, we show examples of how streams might impact the spectrum of the modulation amplitude, as observed by CoGeNT. In the left frame, the results from a simple Maxwellian distribution are shown, for a dark matter mass of 10 GeV and an elastic scattering cross section of $\sigma_n=3 \times 10^{-41}$ cm$^2$. As previously emphasized, the overall normalization of the predicted modulation is a factor of a few smaller than is observed. In the central frame, we add an additional tidal stream of dark matter, with a velocity of 475 km/s, a local density of 0.072 GeV/cm$^3$ (24\% of the density of the smooth halo), and a dispersion of 15 km/s. The characteristics of this stream were chosen to explain CoGeNT's rather high error bar at 2.5 keVee (and, for the appropriate choice of the sodium quenching factor, the peak in the DAMA/LIBRA spectrum as well; see the upper right frame of Fig.~\ref{fig:damaAtCogent}). In the right frame, we add a second stream, with a velocity of 165 km/s, a local density of 0.045 GeV/cm$^3$ (15\% of the density of the smooth halo), and a dispersion of 15 km/s.

Given the presently limited resolution of numerical simulations, it is difficult to assess the probability of significant tidal streams being present in our local neighborhood. Relatively small streams many order of magnitude below the length scales that can be currently resolved could be very important. As an approximate lower limit, we note that current simulations~\cite{simulations} find significant streams to be present at our location of the Milky Way in roughly a few percent of realizations~\cite{Kuhlen:2009vh}.

\subsection{Velocity-Dependent Dark Matter Scattering}
\label{idm}

If the dark matter's scattering cross section with nuclei increases with the velocity of the dark matter particle, the degree of seasonal variation in the observed rate can be larger than is predicted in the standard (velocity-independent) case. Inelastic dark matter scenarios are a well known example of models in which the dark matter possesses velocity-dependent cross sections. In such models, the dark matter can only scatter with nuclei by being excited into a slightly heavier (typically on the order of 100 keV) state~\cite{inelastic}. This requirement suppresses the rate of low energy events, and can increase the degree of annual modulation. 

Inelastic dark matter, however, does not appear help in reconciling the spectra observed by CoGeNT and CRESST-II with the modulation of DAMA/LIBRA and CoGeNT. In particular, the spectrum of events from inelastically scattering dark matter is predicted to be quite flat at low energies, unlike that observed by CoGeNT and CRESST-II.

Other models which introduce a velocity dependent scattering cross section include form factor dark matter~\cite{momentumdm} and resonant dark matter models~\cite{resonant} (see also Refs.~\cite{Fitzpatrick:2010br,Farina:2011pw}). Each of these classes of models hold promise for potentially explaining the large degree of modulation observed by DAMA/LIBRA and CoGeNT. In the first case, a new form factor is introduced which induces a momentum dependence in the interaction between dark matter and nuclei, enhancing the cross section for higher velocity dark matter particles. While this feature can boost the observed modulation amplitude, it will also distort the spectrum of events. In the case of resonant dark matter, the interaction cross section is significantly enhanced near a particular center-of-mass energy, leading to large (and potentially narrow in energy) modulation amplitudes. 

As the CoGeNT collaboration collects more data, the spectrum of the modulation amplitude will become rapidly better measured, making it possible to begin to discriminate between the various options described in this section. By the summer of 2012, CoGeNT will have doubled the size of its data set, and with less background contamination from L-shell electron capture peaks than was present in earlier data. In addition, the CoGeNT collaboration plans to deploy the first of four CoGeNT-4 (C4) detectors in early 2012, roughly quadrupling their effective target mass. If streams or resonances are responsible for a significant fracton of the observed modulation, these features will become increasingly apparent as this data set grows.


\section{Summary and Discussion}
\label{conclusion}

In this article, we have compared the signals reported by the DAMA/LIBRA, CoGeNT, and CRESST-II collaborations. We summarize our finding as follows:
\begin{itemize}

\item{The spectra of events reported by CoGeNT and CRESST-II are in good agreement, and (for a typical Maxwellian velocity distribution) are consistent with a dark matter particle with a mass of approximately 10-20 GeV and an elastic scattering cross section of $\sigma_{n}\approx (1-3)\times 10^{-41}$ cm$^2$. This range of parameter space is roughly consistent with the constraints from CDMS-II, but is in tension with the constraints of xenon-based experiments unless the response of liquid xenon to very low-energy nuclear recoils is lower than previously claimed, or the dark matter's couplings to protons and neutrons destructively interfere for a xenon target.}

\item{The spectra of the modulation amplitudes reported by DAMA/LIBRA and CoGeNT are also consistent with each other. Under the assumption of a typical Maxwellian velocity distribution, these modulation signals favor dark matter particles with masses of 8-19 GeV and an elastic scattering cross section of $\sigma_{n}\approx (0.7-3)\times 10^{-40}$ cm$^2$.}

\item{The apparent mismatch between the elastic scattering cross sections required to produce the event spectra observed by CoGeNT and CRESST-II and those needed to produced the modulations reported by DAMA/LIBRA and CoGeNT could be potentially resolved if the local dark matter distribution contains streams or other highly non-Maxwellian features, or if the dark matter's scattering cross section with nuclei is velocity-dependent.}

\end{itemize}

Taken together, these data appear to favor a dark matter particle with a mass of approximately 10-15 GeV, and an elastic scattering cross section of roughly $\sigma_n \sim 2\times 10^{-41}$ cm$^2$. This mass range is of particular interest in light of recent indirect detection results. In particular, the spatial morphology and spectrum of gamma-rays observed from the Galactic Center can be explained by the annihilations of a 7-12 GeV dark matter particle, annihilating primarily to leptons, and with an annihilation cross section approximately equal to the value required to generate the observed cosmological abundance of dark matter in the early Universe ($\sigma v\sim 3 \times 10^{-26}$ cm$^3$/s)~\cite{GC2,GC1}. The same dark matter model (mass, annihilation cross section, annihilation channels, and halo profile) has also been shown to lead to the production of a diffuse haze of synchrotron emission consistent with that observed by WMAP~\cite{haze}. It also appears that the excess radio emission observed at higher galactic longitudes by the ARCADE 2 experiment~\cite{arcade} possesses a spectral shape and overall intensity consistent with originating from dark matter with the same mass, cross section, dominant channels, and distribution~\cite{arcadedarkmatter,pc}. Lastly, we mention that $\sim$10 GeV dark matter particles with the same distribution and annihilation cross section would be capable of depositing the required energetic electrons into the Milky Way's non-thermal radio filaments~\cite{filaments}, providing an explanation for their peculiar spectral features. Comparing these results to the CRESST-II, CoGeNT, and DAMA/LIBRA signals discussed in this paper, it may be the case that these experiments are each observing different facets of the same species of dark matter.

\bigskip
\bigskip

{\it Acknowledgements}: While in the final stages of this project, Ref.~\cite{kopp} appeared on the arXiv, which addresses many of the issues discussed in this paper. The authors are supported by the US Department of Energy. DH is also supported by NASA grant NAG5-10842. CK is supported by a Fermilab Fellowship in Theoretical Physics.

\end{document}